%% file: M_ToV-final.tex
\documentclass[paper,english,aps, nofootinbib, prl, twocolumn, lengthcheck, amsmath,amssymb,floatfix,showpacs, superscriptaddress,showkeys]{revtex4-1}

\usepackage{booktabs}
\usepackage[T1]{fontenc}
\usepackage[colorlinks=true,linktocpage=true,linkcolor=blue,citecolor=blue]{hyperref}
\usepackage{color,txfonts}
\usepackage{bm}
\usepackage{amssymb}
\usepackage{graphicx}
\usepackage{slashed}
\usepackage{wrapfig}
\usepackage{bbm}
\usepackage[caption=false]{subfig}
\usepackage{verbatim}
\usepackage{longtable}
\usepackage{captcont}
\usepackage{xcolor}
\usepackage{hyperref}

\usepackage{mathtools}
\usepackage{natbib}
\usepackage{ulem}
\input{aas_macros}

\begin{document}

\title{Maximum gravitational mass $M_{\rm TOV}=2.25^{+0.08}_{-0.07}M_\odot$ inferred at about $3\%$ precision with multimessenger data of neutron stars}

\author{Yi-Zhong Fan}
\email{Corresponding author: yzfan@pmo.ac.cn}
\affiliation{Key Laboratory of Dark Matter and Space Astronomy, Purple Mountain Observatory, Chinese Academy of Sciences, Nanjing 210023, China}
\affiliation{School of Astronomy and Space Science, University of Science and Technology of China, Hefei, Anhui 230026, China}

\author{Ming-Zhe Han}
\email{Corresponding author: hanmz@pmo.ac.cn}
\affiliation{Key Laboratory of Dark Matter and Space Astronomy, Purple Mountain Observatory, Chinese Academy of Sciences, Nanjing 210023, China}

\author{Jin-Liang Jiang}
\email{Corresponding author: jiang@itp.uni-frankfurt.de}
\affiliation{Institut f\"{u}r Theoretische Physik, Goethe Universit\"{a}t, Max-von-Laue-Str. 1, D-60438 Frankfurt am Main, Germany.}

\author{Dong-Sheng Shao}
\affiliation{Key Laboratory of Dark Matter and Space Astronomy, Purple Mountain Observatory, Chinese Academy of Sciences, Nanjing 210023, China}

\author{Shao-Peng Tang}
\email{Corresponding author: tangsp@pmo.ac.cn}
\affiliation{Key Laboratory of Dark Matter and Space Astronomy, Purple Mountain Observatory, Chinese Academy of Sciences, Nanjing 210023, China}

\begin{abstract}
The maximal gravitational mass of nonrotating neutron stars ($M_{\rm TOV}$) is one of the key parameters of compact objects and only loose bounds can be set based on the first principle. With reliable measurements of the masses and/or radii of the neutron stars, $M_{\rm TOV}$ can be robustly  inferred from either the mass distribution of these objects or the reconstruction of the equation of state (EoS) of the very dense matter. 
For the first time we take the advantages of both two approaches to have a precise inference of $M_{\rm TOV}=2.25^{+0.08}_{-0.07}~M_\odot$ (68.3\% credibility), with the updated neutron star mass measurement sample, the mass-tidal deformability data of GW170817, the mass-radius data of PSR J0030+0451 and PSR J0740+6620, as well as the theoretical information from the chiral effective theory 
($\chi$EFT) and perturbative quantum chromodynamics (pQCD) at low and very high energy densities, respectively. This narrow credible range is benefited from the suppression of the high $M_{\rm TOV}$ by the pQCD constraint and the exclusion of the low $M_{\rm TOV}$ by the mass function.
Three different EoS reconstruction methods are adopted separately, and the resulting $M_{\rm TOV}$ and $R_{\rm TOV}$ are found to be almost identical, where $R_{\rm TOV}=11.90^{+0.63}_{-0.60}$ km is the radius of the most massive non-rotating NS. This precisely evaluated $M_{\rm TOV}$ suggests that the EoS of neutron star matter is just moderately stiff and the $\sim 2.5-3M_\odot$ compact objects detected by the second generation gravitational wave detectors are most likely the lightest black holes.    
\end{abstract}

\maketitle

\section{Introduction}  
For a given equation of state (EoS) of the ultra-dense matter, the maximum gravitational mass of a nonrotating neutron star $M_{\rm TOV}$ is uniquely determined by the well-known Tolman-Oppenheimer-Volkoff equations \cite{PhysRev.55.374}. A $M_{\rm TOV}$, if known experimentally or theoretically, would in turn play a  crucial role in reconstructing the EoS of the ultra-dense matter \cite{2012ARNPS..62..485L}. However, the actual value of $M_{\rm TOV}$ remains uncertain. One widely-adopted theoretical upper limit, which bases  on the general relativity, the principle of causality, and Le Chatelier’s principle, is  $M_{\rm TOV} \leq  2.9 M_\odot$ \cite{PhysRevLett.32.324,1996ApJ...470L..61K}. Observationally, the most massive neutron star (NS) measured so far imposes a lower limit on $M_{\rm TOV}$. So far, among the accurate measurements, the record is held by PSR J0740+6620 with a mass of $2.08\pm 0.07M_\odot$ \cite{Fonseca_2021}. Throughout this work, the error bars are for 68.3\% confidence level unless specifically mentioned. 

One interesting possibility to infer the $M_{\rm TOV}$ is via the modeling of short gamma-ray bursts (GRBs) that have been widely believed to be formed by double NS mergers \cite{1989Natur.340..126E}. A $M_{\rm TOV}\sim 2.3M_\odot$ is preferred in the supramassive neutron star model for some short GRBs with a peculiar X-ray plateau \cite{2013PhRvD..88f7304F}. Assuming a black hole or a long-lived NS central engine for short GRBs, $M_{\rm TOV}\sim 2.2M_\odot$ or $\sim 2.5M_\odot$ is needed \cite{2015ApJ...808..186L}. After the discovery of GW170817/GRB 170817A \cite{LIGOScientific:2017vwq,Goldstein:2017mmi}, the application of the above ideas have yielded a constraint of $M_{\rm TOV} \leq 2.2-2.3M_\odot$ (e.g., \cite{Margalit:2017dij,PhysRevD.96.123012,Rezzolla:2017aly,Ma:2017yva,PhysRevD.97.021501}). A $M_{\rm TOV}=2.17\pm 0.09M_\odot$ has been inferred \cite{2020ApJ...904..119F,2020PhRvD.101f3029S} via modeling the multi-messenger data of GW 170817/GRB 170817A/AT 2017gfo \cite{LIGOScientific:2017vwq,Goldstein:2017mmi,LIGOScientific:2017ync}. Though attractive, such a result relies on two {\it strong} assumptions: the central engine of GRB 170817A was a black hole formed in the collapse of a supramassive NS at $t\sim  0.8$ s after the merger, and the so-called universal relationships among the parameters of the cold NSs are still applicable to the nascent remnants formed in the NS mergers with a temperature of tens of MeV (the second uncertainty also applies to other upper bounds on $M_{\rm TOV}$ set with the data of GW170817). 

A reliable inference of $M_{\rm TOV}$, which does not suffer from the above uncertainties, can be achieved in 
reconstructing the EoS of the ultra-dense matter with the multi-messenger data of NSs. This has been extensively investigated in the literature but the uncertainties are still high \cite[e.g.,][]{PhysRevD.101.123007,PhysRevC.102.055803,PhysRevD.104.063032,2020ApJ...892...55J}. 
The perturbative quantum chromodynamics (pQCD) inputs impact the inference of the neutron-star-matter EoS \cite{2020NatPh..16..907A,2022PhRvL.128t2701K} and can effectively narrow down the posterior distribution of $M_{\rm TOV}$  \cite{Gorda:2022jvk,Han:2022rug}. Very recently, a Bayesian inference of the EoS via a single-layer feed-forward neural network (FFNN) yields $M_{\rm TOV}=2.18^{+0.27}_{-0.13}M_\odot$ (90\% credibility; \cite{Han:2022rug}), where just the direct mass/radius measurements of GW170817, PSR J0030+0451, and PSR J0740+6620 as well as the theoretical constraints from 
chiral effective theory ($\chi$EFT) and pQCD have been taken into account and 
hence does not suffer from serious observational bias. 
Anyhow, the credible range of the resulting $M_{\rm TOV}$ is still relatively broad and a considerable improvement will be essential for distinguishing the EoSs proposed in the literature. It is also crucial to check whether the bounding on $M_{\rm TOV}$ is sensitively dependent of the (non)parameterization methods of EoS and the density to impose the pQCD constraint ($n_{\rm L}$) or not. Moreover, the growing population of the NSs with reliable mass measurements
makes it feasible to statistically infer the features of the distribution, and a cutoff at the high end of the mass distribution ($M_{\rm max}$) can be reasonably interpreted as $M_{\rm TOV}$ \cite{Alsing:2017bbc,Shao:2020bzt, 2020PhRvD.102f4063C, 2023Galax..11...19D}. 

In order to get a precise and model-insensitive value of $M_{\rm TOV}$ we have the following {\it novel} treatments: (i) The maximum cutoff of the mass distribution of the NSs measured so far is taken as a likelihood function of $M_{\rm TOV}$ in the analysis to narrow down the parameter space;  (ii) The EoS of the NS matter is reconstructed with three very different (non)parameterization models; (iii) The choice/influence of $n_{\rm L}$ is examined. Our main finding is the  precision inference of $M_{\rm TOV}=2.25^{+0.08}_{-0.07}~M_\odot$ and $R_{\rm TOV}=11.90^{+0.63}_{-0.60}$ km. 

\section{Methods} 
\label{sec:methods}
Benefiting from the progresses made on
pulsar radio timing and X-ray observations, the growing
population of the neutron stars with reliable mass measurements
within decades makes it feasible to statistically infer the
features of the mass distribution. A cutoff at the
high end of the mass distribution, most-likely represents $M_{\rm TOV}$, has been reported in the literature \cite{Alsing:2017bbc,Shao:2020bzt}. 
In this work we simultaneously constrain the NS mass distribution and the maximum mass cutoff with the latest sample (see Table 1 in the Appendix). We infer the mass distribution parameters and the maximum mass cutoff using the hierarchical Bayesian inference \cite{Thrane:2018qnx} with the following likelihood
\begin{equation}
    \mathcal{L}(\boldsymbol{d}|\vec{\theta}) \propto \prod_{i=1}^{N} \left(\frac{1}{n_i}\sum_{k}^{n_i} P(m_i^k|\vec{\theta})\right),
\end{equation}
where $N$ is the size of NS sample, $P(m_i^k|\vec{\theta})$ is a two-component Gaussian mixture mass distribution model parameterized with $\vec{\theta}$ (see Eq.(1) in \cite{Shao:2020bzt}), and $m_i^k$ is the $k$-th sample of the $i$-th mass measurement.
The samples for each mass measurement are randomly generated by approximating the reported error bar with an asymmetric normal distribution \cite{Kiziltan:2013oja}. The priors of mass distribution parameters are uniform with ranges identical to those used in \cite{Shao:2020bzt}.
We have further examined the effects of employing a Gaussian+Cauchy-Lorentz distribution model for $P(m_i^k|\vec{\theta})$ and the exclusion of certain high-mass NSs and the results are found to be similar, as elaborated in the Appendix.
The resulting maximum mass cutoff distribution is taken as an additional likelihood in our further EoS reconstruction.

The data set included in our EoS inference is the same as that of \cite{Han:2022rug} since there are no new direct simultaneous mass/radius measurements of the NSs from either LIGO/Virgo/KAGRA network or Neutron Star Interior Composition Explorer (NICER).
Again, in the low density regime, we match the constructed EoS to the NS crust EoS \cite{Douchin:2001sv} down to $0.3n_{\rm s}$, where $n_{\rm s}$ denotes the nuclear saturation number density.
At higher densities, the EoS is modelled in three different ways to proceed the Bayesian inference.
The high precision (i.e., the next-to-next-to-next-to leading order, $\rm N^3LO$) $\chi$EFT calculation results \cite{Drischler:2017wtt} serve as the constraint, i.e., we exclude the EoS that exceeds the 3$\sigma$ range of $\chi$EFT for the FFNN and flexible piecewise linear sound speed (PWLS) methods \citep{Jiang:2022tps}, while for GP method we take it as the training data for conditioning the GP process (see bellow for more information about these three methods).
At the very high density regime (i.e., $\geq 40n_{\rm s}$), the pQCD calculation provides reliable information. The corresponding causality-driven constraint from pQCD can be pushed to a considerably lower density $n_{\rm L}$ \cite{2022PhRvL.128t2701K}. One widely-adopted approach is to extrapolate the $\chi$EFT and NS data bounded EoS to $n_{\rm L}\geq 10n_{\rm s}$ and then reject the models violating the pQCD constraints \cite[e.g.,][]{2020NatPh..16..907A,2022PhRvL.128t2701K,Han:2022rug,Jiang:2022tps,Ecker:2022xxj,Sinan2022}. Such an approach yields a very soft core in the most massive NSs, indicating the presence of quark matter \cite{2020NatPh..16..907A,Han:2022rug}. The impact of pQCD constraints on the EoS properties, in particular in the core of the most massive NSs, is found to depend sensitively on the choice of $n_{\rm L}$ \cite{Somasundaram:2022ztm,Essick:2023fso}. This is anticipated since the pQCD constraints were extrapolated into low densities from $n \geq 40n_{\rm s}$ with some general conditions and become much less constrainable at central density $n_{\rm c,TOV}\sim 5n_{\rm s}$ (see Fig.~1 of \cite{2022PhRvL.128t2701K} for demonstration) of non-rotating NS with maximum mass. We thus still adopt $n_{\rm L}=10n_{\rm s}$ for a fiducial choice, as validated in Sec.~\ref{sec:nLchoice} (Taking $n_{\rm L}= n_{\rm c,TOV}$ will just moderately broadens the uncertainty range of  $M_{\rm TOV}$ to $\approx 0.1M_\odot$).
 
Once the EoS is constructed, the relations between the macroscopic properties of NS is predicted and can be used to carry out the Bayesian inference for the observation data to obtain the posterior distributions of the EoS. The overall likelihood of the Bayesian inference, which is rather similar to that of \cite{Han:2022rug}, is expressed as
\begin{eqnarray}
    \mathcal{L} &= \mathcal{L}_{\rm GW}
    \times \mathcal{L}_{\rm NICER}
    \times \mathcal{L}_{\rm pQCD}
    \times \mathcal{L}_{\rm M_{\rm max}}.
\end{eqnarray}
where $\mathcal{L}_{\rm GW}=\mathcal{P}(m_1, m_2, \Lambda_1(m_1, \theta_{\rm EoS}), \Lambda_2(m_1, \theta_{\rm EoS}))$ is the marginalized likelihood of the GW170817 \citep{2020MNRAS.499.5972H} ($m_{1, 2}$ and $\Lambda_{1, 2}$ are the mass and tidal deformability of the primary/secondary NS in GW170817), and $\theta_{\rm EoS}$ is the set of parameters used to determine the EoS; $\mathcal{L}_{\rm NICER}=\prod_i \mathcal{P}_i(M(\theta_{\rm EoS}, h_i), R(\theta_{\rm EoS}, h_i))$ is the likelihood of the two NICER observations ($M$, $R$, and $h$ are the mass, radius, and core pseudo enthalpy of the NS, respectively),
and we use the Gaussian kernel density estimation (KDE) of the public posterior samples of the data from two observations, PSR J0030+0451 \citep{2019ApJ...887L..21R} and PSR J0740+6620 \citep{2021ApJ...918L..27R};
$\mathcal{L}_{\rm pQCD}=\mathcal{P}(n_{\rm L}, \varepsilon(n_{\rm L}, \theta_{\rm EoS}), p(n_{\rm L}, \theta_{\rm EoS}))$ is the likelihood of the pQCD constraints implemented at $n_{\rm L} \sim 10n_{\rm s}$ \citep{2022PhRvL.128t2701K}, where $\varepsilon$ and $p$ are the energy density and the pressure, respectively;
{$\mathcal{L}_{\rm M_{\rm max}}=\mathcal{P}(M_{\rm TOV}(\theta_{\rm EoS}))$ is the marginalized posterior distribution of the NS maximum mass cutoff adopted from the Bayesian analysis of NS mass distribution.}

To check the model-dependence of our result, the EoS inference is performed in three different ways. {\it The first} is following \cite{Han:2022rug}, where the neutron star EoS is represented by the 10-node (N=10) single-layer FFNN expansion that is capable of fitting theoretical EoSs very well \citep{2021ApJ...919...11H}, which reads 
\begin{equation}
    c_{\rm s}^2(\rho) = S(\sum_i^{\rm N} w_{2i}\sigma(w_{1i}\ln\rho+b_{1i})+b_{2}),
    \label{eq:Han2023}
\end{equation}
where $c_{\rm s}^2={\rm d}p/{\rm d}\varepsilon$ is the squared sound speed, $\sigma(\cdot)$ is the activation function, $w_{\rm 1i}$, $w_{\rm 2i}$, $b_{\rm 1i}$, and $b_2$ are weights/bias parameters of the FFNN.
$\rho$ is the rest-mass density (to obtain the baryon number density, the average mass of a baryon is taken to be $1.66\times 10^{-24}$ g).
$S(x)=1/(1+e^{-x})$ ranges from 0 to 1, guaranteeing the microscopical stability and causality condition.
We consider two types of activation functions, including the {\it sigmoid} of $1/(1+e^{-x})$ and the {\it hyperbolic tangent} of ${(e^x-e^{-x})}/{(e^x+e^{-x})}$.
The model with ${\it sigmoid}$ prefers the monotonically increasing sound speed or with a gentle peak, while the model with {\it hyperbolic tangent} is more likely to generate the EoSs with an exotic structure like the vanishing sound speed or a sharp peak. 
Therefore, a combination of these two models considerably enlarges the prior space (note that the weights and biases parameters are uniformly sampled in $(-5, 5)$).
The inferred properties of NSs with each activation function are consistent, we thus combine the two sets of posteriors to obtain the results. This approach is distinguished by its capability of generating some physically-motivated EoSs, such as the hadronic, the first-order phase transition, and the quark-hadron crossover, efficiently.
The EoS with $M_{\rm TOV}$  
beyond $1.4-3~M_\odot$ is discarded during the inference.
{\it The second} is the piecewise linear sound speed (PWLS)
method detailed in \citet{Jiang:2022tps}: the EoS below densities  $0.5n_{\rm s}$ are described by the BPS EoS. Beyond this range, it is divided into 11 segments. The first segment is represented by a single polytrope, while the others are characterized by linear segments of $\mu-c_{\rm s}^{2}$ relations, with parameters $c_{{\rm s},i}^{2}$ at logarithmically separated fixed chemical potential positions, where $\mu$ represents the chemical potential. The likelihood construction follows the same approach as the first method to ensure comparability of results. The pQCD constraints, as described previously, are slightly different from that of \citet{Jiang:2022tps}.
We use the Bayesian inference library {\sc BILBY} \citep{2019ApJS..241...27A} with the sampling algorithm {\sc PyMultiNest} to obtain the posterior samples of the EoSs in these two EoS construction methods.
{\it The third} is the widely used Gaussian Process (GP) regression introduced in \citet{PhysRevD.101.123007} and \citet{PhysRevC.102.055803}. Unlike these two works, we use $\phi$ as a function of the baryon number density $n$ to describe the NS EoS, where the $\phi$ is an auxiliary variable defined as 
\begin{equation}
    \phi = -\ln(1/c_{\rm s}^2 -1),
\end{equation}
which can guarantee the microscopic stability and the causality simultaneously, i.e., $0 \leq c_{\rm s}^2 \leq 1$. A GP can be considered as a multivariate Gaussian distribution with infinite dimensions, and it is controlled by several hyper parameters. The GP of $\phi(n)$ can be described as 
\begin{equation}
    \phi(n) \sim \mathcal{N}(-\ln(1/\bar{c_{\rm s}}^2 -1), K(n_i, n_j)),
\end{equation}
where $K(n_i, n_j)=\eta^2 e^{-(n_i-n_j)^2/2l^2}$ is the kernel function of the GP, and ${\bar{c}_{\rm s}^2, \eta^2, l}$ are the hyper parameters: the mean of squared sound speed, the variance, and the correlation length, respectively. Our choice of the hyper parameters mainly follows \citet{Gorda:2022jvk}, which are randomly drawn from the hyper priors, i.e.,
$\bar{c}_{\rm s}^2 \sim \mathcal{N}(0.5, 0.25^2), \eta^2 \sim \mathcal{N}(1.25, 0.2^2)$, and  
$l \sim \mathcal{N}(0.5n_{\rm s}, (0.25n_{\rm s})^2)$.
We choose the mean of $l$ to be $\bar{l}=0.5n_{\rm s}$ rather than $1n_{\rm s}$ used in \cite{Gorda:2022jvk}, since the current uncertainty of $n_{\rm c,TOV}$ is about $1n_{\rm s}$ and a smaller $\bar{l}$ may be helpful to better trace the possible change of the matter properties at $n \sim n_{\rm c,TOV}$. The EoS posterior of GP method is selected according to their likelihoods (or weights) from the generated sample.

\section{Results} 
\subsection{The $M_{\rm max}$ inferred in the mass function modeling}
\begin{figure}[htbp]
\centering
\vspace{-0.3cm}
\includegraphics[width=0.49\textwidth]{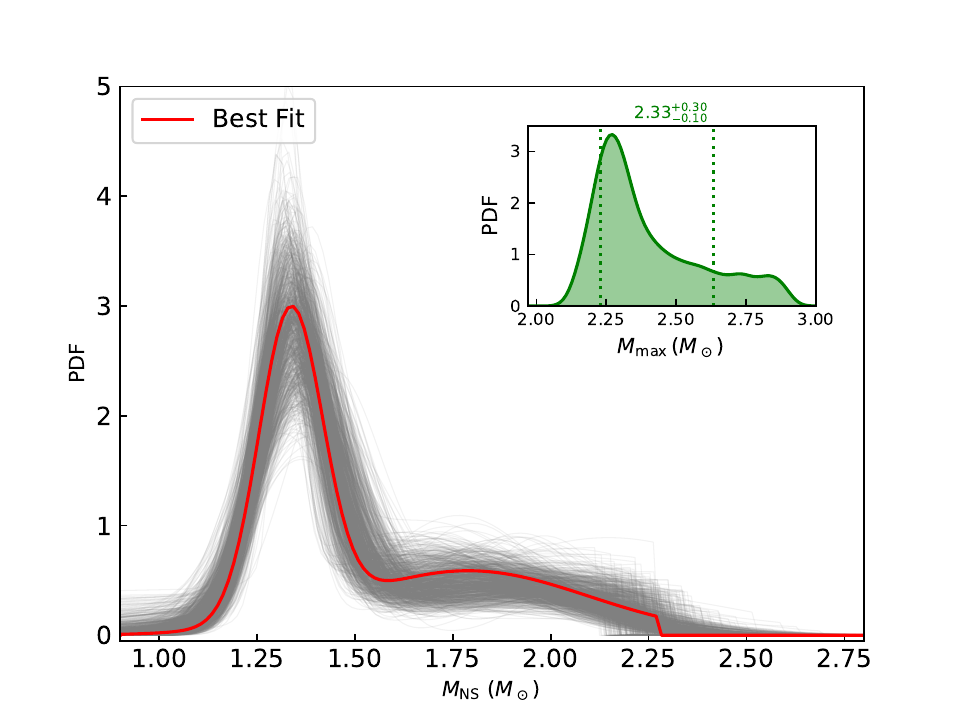}
\vspace{-0.3cm}
\caption{The red line represents the best-fit mass distribution, i.e., a two-component Gaussian mixture with a sharp cutoff of $M_{\rm max}=2.28~M_{\odot}$, of the 136 neutron stars with gravitational mass measurements. Here we take 1000 independent posterior samples (the grey lines) to give a visual guide for the uncertainties. The inset shows $P(M_{\rm max})$, the posterior distribution of $M_{\rm max}$.}
\label{fig:mass_dist}
\end{figure}

In \cite{Shao:2020bzt} we have collected the masses for 103 NSs. Up to April 2023, there are 34 new measurements and 16 out of the previous NS mass sample have been updated with the more accurate data. Therefore, the current NS mass sample summarized in Table 1 of the Appendix consists of 136 objects, including 23 double NS binary systems, 68 NS-white dwarf binary systems, and 23 X-ray binary systems. We adopt the two-component Gaussian mixture model with a maximum mass cutoff $M_{\rm max}$ (see Sec. II B of \cite{Shao:2020bzt} for the details) to fit these NS mass data. We show in Fig.~\ref{fig:mass_dist} the best fit result and a visual impression of the uncertainties on the shape of the distribution with 1000 independent posterior samples found in our modeling. The inset presents $P(M_{\rm max})$,
the posteriori distribution of $M_{\rm max}$, and the 68.3\% and 90\% credible regions are $2.33^{+0.30}_{-0.10}~M_{\odot}$ and $2.33^{+0.49}_{-0.15}~M_{\odot}$ respectively. For the impact of different mass distribution models or NS sample on the $M_{\rm max}$, please see the Appendix, where we find that the mass distribution models have minor influence on the overall results. {The $M_{\rm max}$ obtained in this work is slightly larger than our previous finding $M_{\rm max}\sim2.26M_\odot$ \citep{Shao:2020bzt}, which is primarily attributed to the enlargement of the data sample.}

\subsection{The EoS reconstruction results with different (non)parameterization approaches}
With the radius/mass data of PSR J0030+0451, PSR J0740+6620 and the double NSs involved in GW170817, we have reconstructed the EoS of the dense matter with three (non)parameterization methods summarized in Sec. \ref{sec:methods}, in which the theoretical $\chi$EFT and pQCD ($n_{\rm L}=10n_{\rm s}$) constraints and the information of $P(M_{\rm max})$ have been incorporated.

\begin{figure}[htbp]
\centering
\vspace{-0.3cm}
\includegraphics[width=0.5\textwidth]{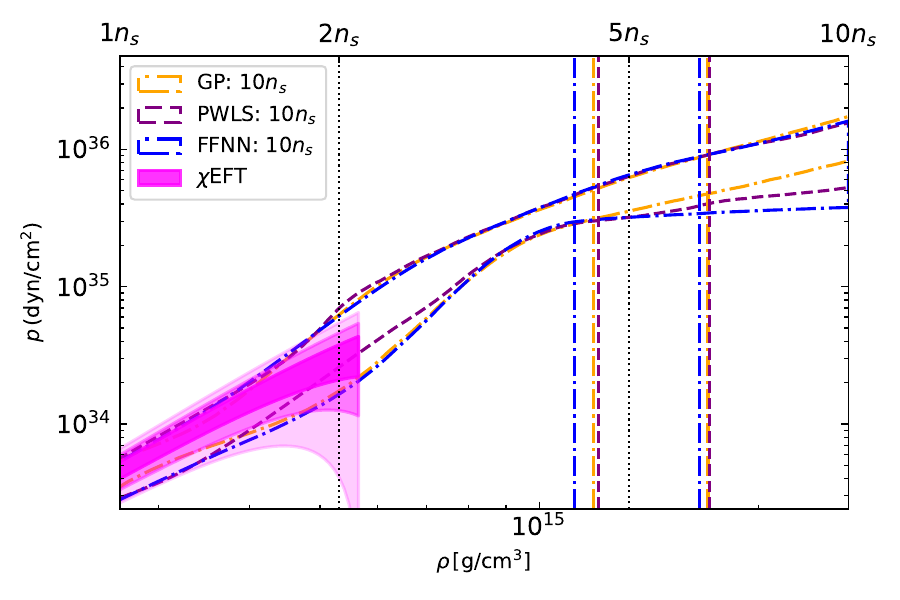}
\vspace{-0.3cm}
\caption{The EoS reconstruction results for the single-layer feed-forward neural network (FFNN) model of \citet{Han:2022rug} (the blue dash-dotted curve), the Gaussian process method mainly following \citet{Gorda:2022jvk} (the orange dash-dotted line), and the flexible piecewise method of \citet{Jiang:2022tps} (the purple dashed curve) in the case of $n_{\rm L}=10n_{\rm s}$.
In the plot the regions are for 90\% confidence level (the vertical lines mark the regions of inferred $n_{\rm c,TOV}$).}
\label{fig:eos_range}
\end{figure}

In Fig.~\ref{fig:eos_range} we present the EoS reconstruction results with three different (non)parameterization methods. The data sets for the three constraints are the same, including the latest calculation results of the $\chi$EFT and pQCD (extrapolated to 10$n_{\rm s}$), the four neutron stars with mass and radius/tidal deformability, and the information on the possibility of $M_{\rm TOV}$ inferred from the mass distribution of the neutron stars with measured gravitational masses. The 90\% credible regions are similar, in particular in the density region of $\leq n_{\rm c,TOV}$, which can well explain the almost identical $M_{\rm TOV}$ as well as $R_{\rm TOV}$ shown in Fig.~\ref{fig:M_TOV}.  

\subsection{The constraints on $M_{\rm TOV}$ and the role of the choice of $n_{\rm L}$}\label{sec:nLchoice}

\begin{figure}[htbp]
\centering
\vspace{-0.3cm}
\includegraphics[width=0.5\textwidth]{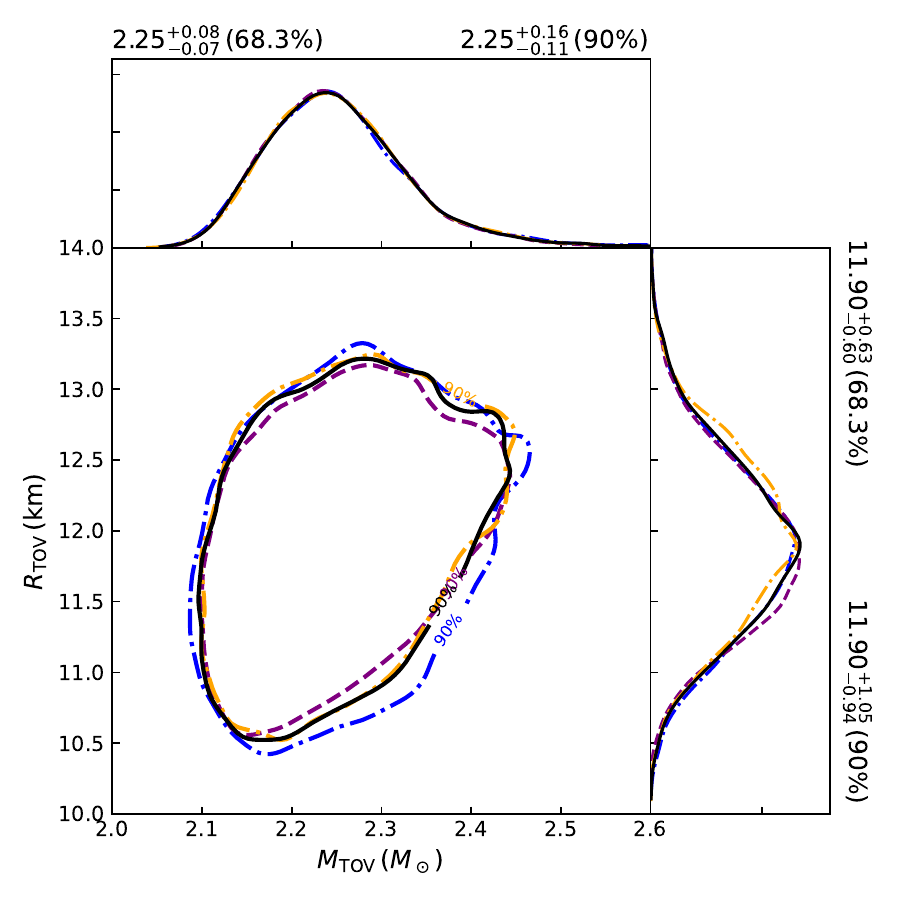}
\vspace{-0.3cm}
\caption{Almost identical posterior distributions of $M_{\rm TOV}$ and $R_{\rm TOV}$ found in three independent EoS reconstruction approaches for $n_{\rm L}=10n_{\rm s}$, where we have taken into the mass/radius data of four NSs (including PSR J0030+0451, PSR J0740+6620 and the two involved in GW170817), the posterior distribution of the maximum cutoff of the neutron star mass function, as well as the theoretical $\chi$EFT and pQCD constraints. The orange, purple and blue lines represent the results for the GP, PWLS, and FFNN models, while the black gives their average.} 
\label{fig:M_TOV}
\end{figure}

Here we focus on the constraint on $M_{\rm TOV}$. Fig.~\ref{fig:M_TOV} shows the posterior distributions of $M_{\rm TOV}$ and $R_{\rm TOV}$ found in three independent EoS reconstruction approaches in different colors. These approaches yield almost identical results on both $M_{\rm TOV}$ (i.e., $2.25^{+0.08}_{-0.07}M_\odot$) and $R_{\rm TOV}$ (i.e., $11.90^{+0.63}_{-0.60}$ km), i.e., they are model-insensitive. At 95\% confidence level we have $M_{\rm TOV}\leq 2.4M_\odot$, which is well consistent with that found in a semi-analytical study \cite{2019EPJA...55...39Z}.

\begin{figure}[htbp]
\centering
\vspace{-0.3cm}
\includegraphics[width=0.5\textwidth]{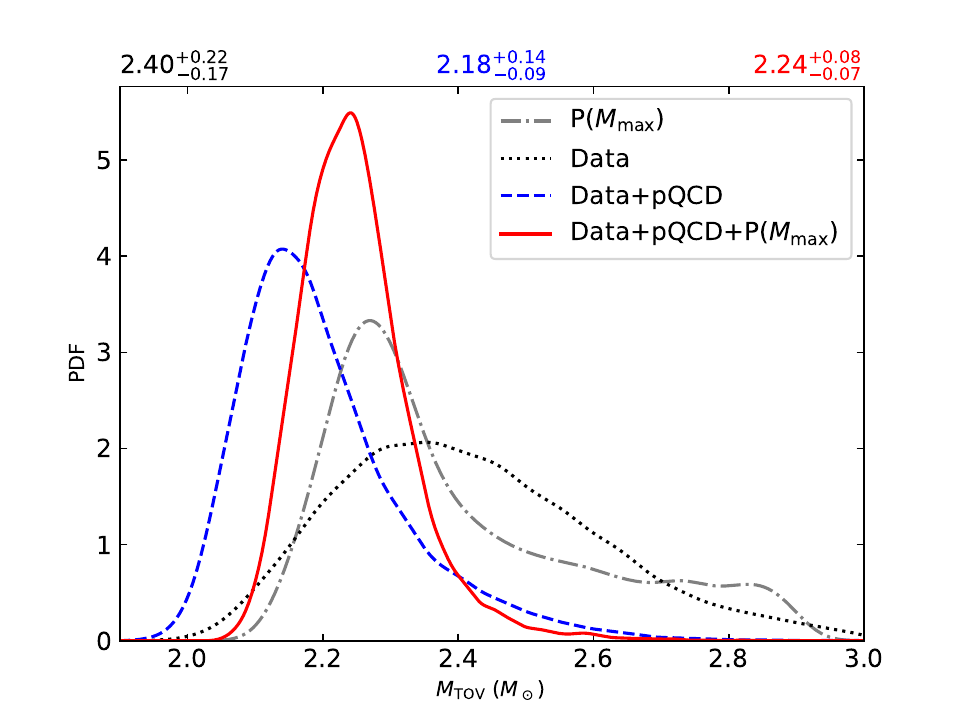}
\vspace{-0.3cm}
\caption{$M_{\rm TOV}$ obtained with different data sets and/or conditions. The grey dash-dotted line represents $P(M_{\rm max})$. The black dotted line represents that deduced from the $\chi$EFT constraint as well as the four NSs with both mass and radius measurements. The blue dashed curve employs also the pQCD constraint. The red solid line further takes into account the information of the mass distribution of NSs. The last three lines are calculated with the FFNN approach.}
\label{fig:mtov_dist}
\end{figure}

In Fig.~\ref{fig:mtov_dist} we show the roles of different sets of data or conditions in narrowing down the range of $M_{\rm TOV}$. The blue dashed curve, characterized by a broad distribution (i.e., it is loosely constrained), represents the posterior of $M_{\rm TOV}$ bounded with the mass/radius of the four NSs as well as the theoretical $\chi$EFT constraint. Thanks to the additional input of the pQCD likelihood, the posterior of $M_{\rm TOV}$ gets effectively suppressed in the high mass range (see the black dotted curve), in agreement with that found in \cite{Gorda:2022jvk,Han:2022rug}. The further inclusion of the $P(M_{\rm max})$ likelihood in the constraint narrows the allowed low mass range of $M_{\rm TOV}$, which is anticipated since $P(M_{\rm max})$ concentrates in the relatively high mass range. Consequently, the $M_{\rm TOV}$ is precisely inferred  (see the red line in Fig.~\ref{fig:mtov_dist}).

\begin{figure}[htbp]
\centering
\vspace{-0.3cm}
\includegraphics[width=0.5\textwidth]{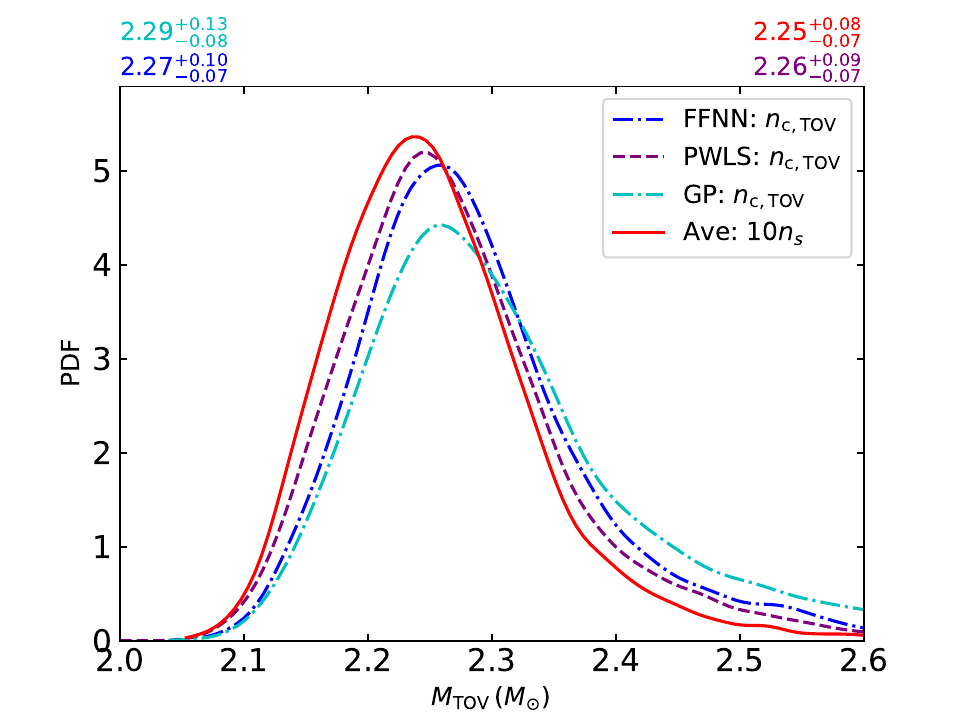}
\includegraphics[width=0.5\textwidth]{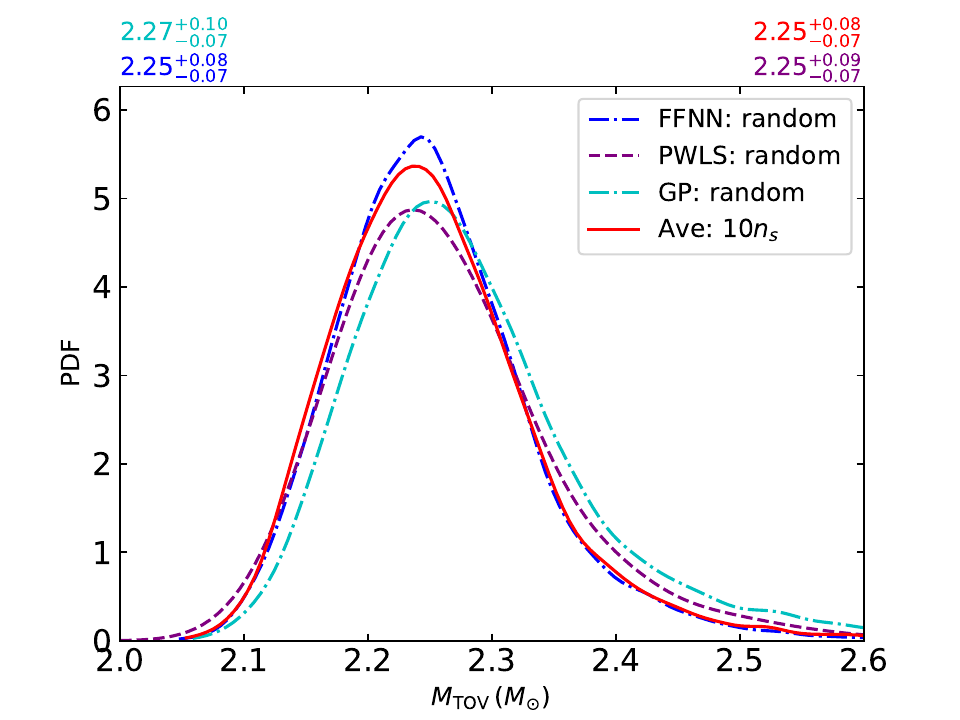}
\includegraphics[width=0.5\textwidth]{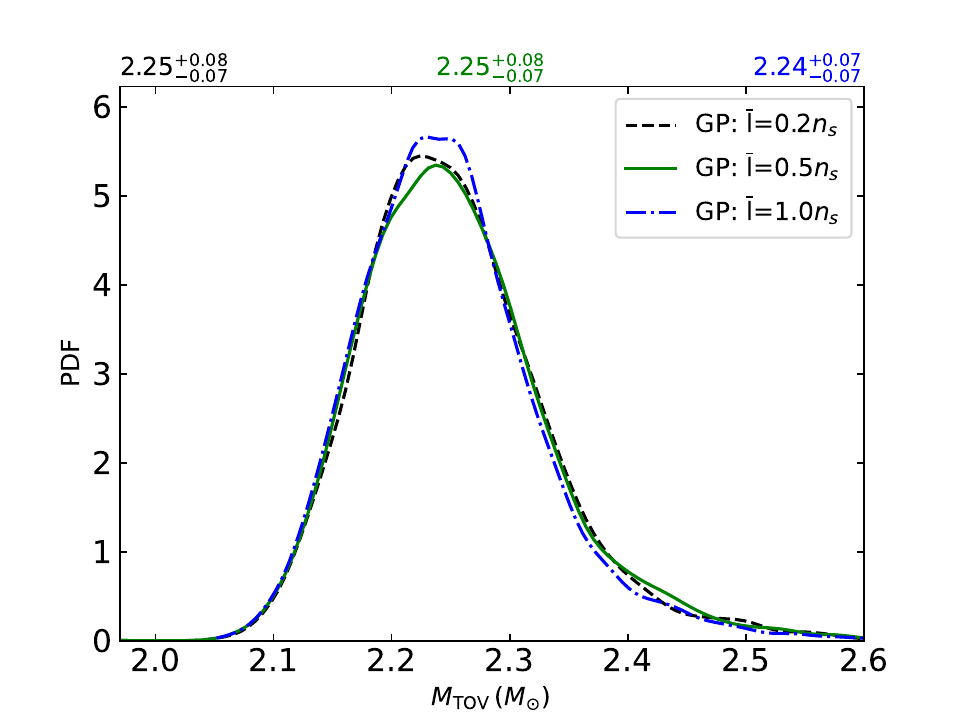}
\vspace{-0.3cm}
\caption{The top panel: the posterior distributions of the $M_{\rm TOV}$ found in the three EoS reconstruction approaches in the case of $n_{\rm L}=n_{\rm c,TOV}$, which are moderately weaker than the case of $n_{\rm L}=10n_{\rm s}$ (i.e., the red line). 
The middle panel: the posterior distribution of $M_{\rm TOV}$ obtained in setting the pQCD constraint at a density uniformly distributed between $\rho_{\rm TOV}$ and $10n_{\rm s}$ (i.e, the random check). 
The bottom panel: the posterior distribution of $M_{\rm TOV}$ obtained in the GP model with different $\bar{l}$.}
\label{fig:M_TOV-check}
\end{figure}

It has been pointed out that the impact of pQCD constraints on the EoS properties, in particular in the core of the most massive NSs, depends sensitively on the choice of $n_{\rm L}$ \cite{Somasundaram:2022ztm,Essick:2023fso}. This is because that the pQCD constraints were extrapolated into low energy densities from $n \geq 40n_{\rm s}$ with some general conditions and become much less constrainable at $n_{\rm c,TOV} \sim 5n_{\rm s}$. Though we think it is reasonable to take $n_{\rm L}=10n_{\rm s}$ to get tight constraints, as a check here we also take $n_{\rm L}=n_{\rm c,TOV}$ to infer $M_{\rm TOV}$. Again, the three EoS (non)parameterization methods mentioned in Sec. \ref{sec:methods} are adopted. The results are shown in Fig.~\ref{fig:M_TOV-check}. In general, the resulting $M_{\rm TOV}$ are still well consistent with each other and we do not see strong model dependence. 
We also present the averaged posterior (red curve in both top and middle panels of Fig.~\ref{fig:M_TOV-check}) resulting from three different methods when applying pQCD constraints at $n_{\rm L}=10n_{\rm s}$. 
Now we find a bit higher median value of $M_{\rm TOV}$ and a moderately broadened error bar of $\approx 0.1M_\odot$. This broadening can be understood since the pQCD constraint at $n_{\rm L}=n_{\rm c,TOV}$ is weaker than that at $n_{\rm L}=10n_{\rm s}$.

\begin{figure}[htbp]
\centering
\vspace{-0.3cm}
\includegraphics[width=0.5\textwidth]{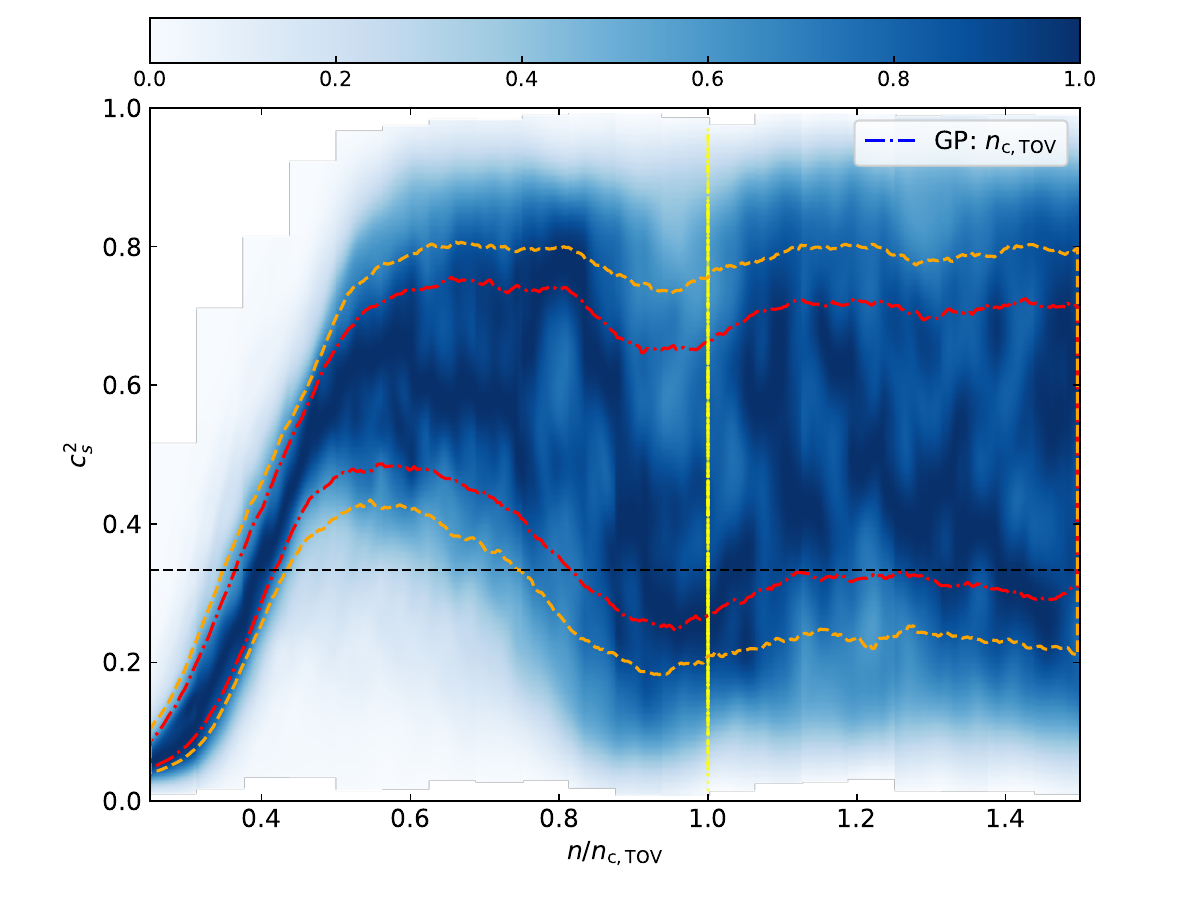}
\includegraphics[width=0.5\textwidth]{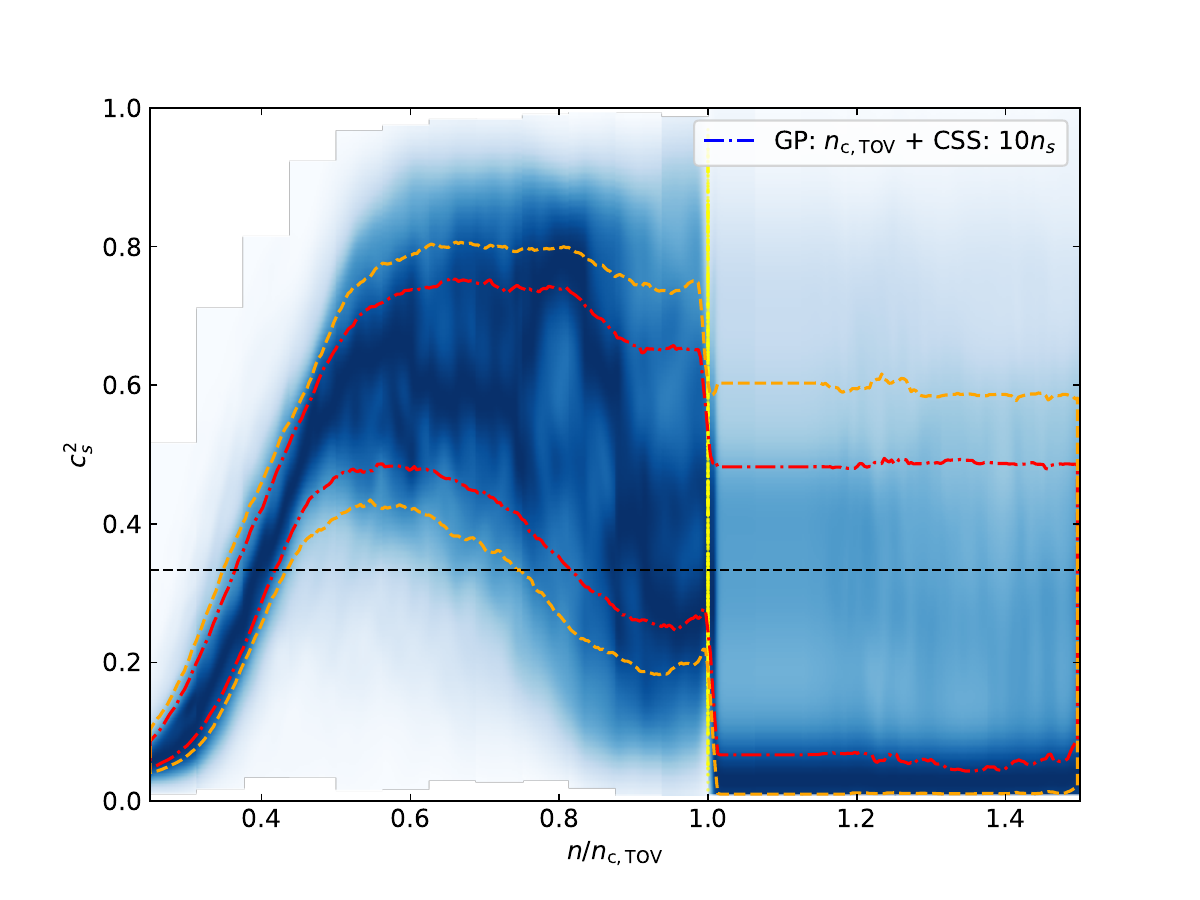}
\includegraphics[width=0.5\textwidth]{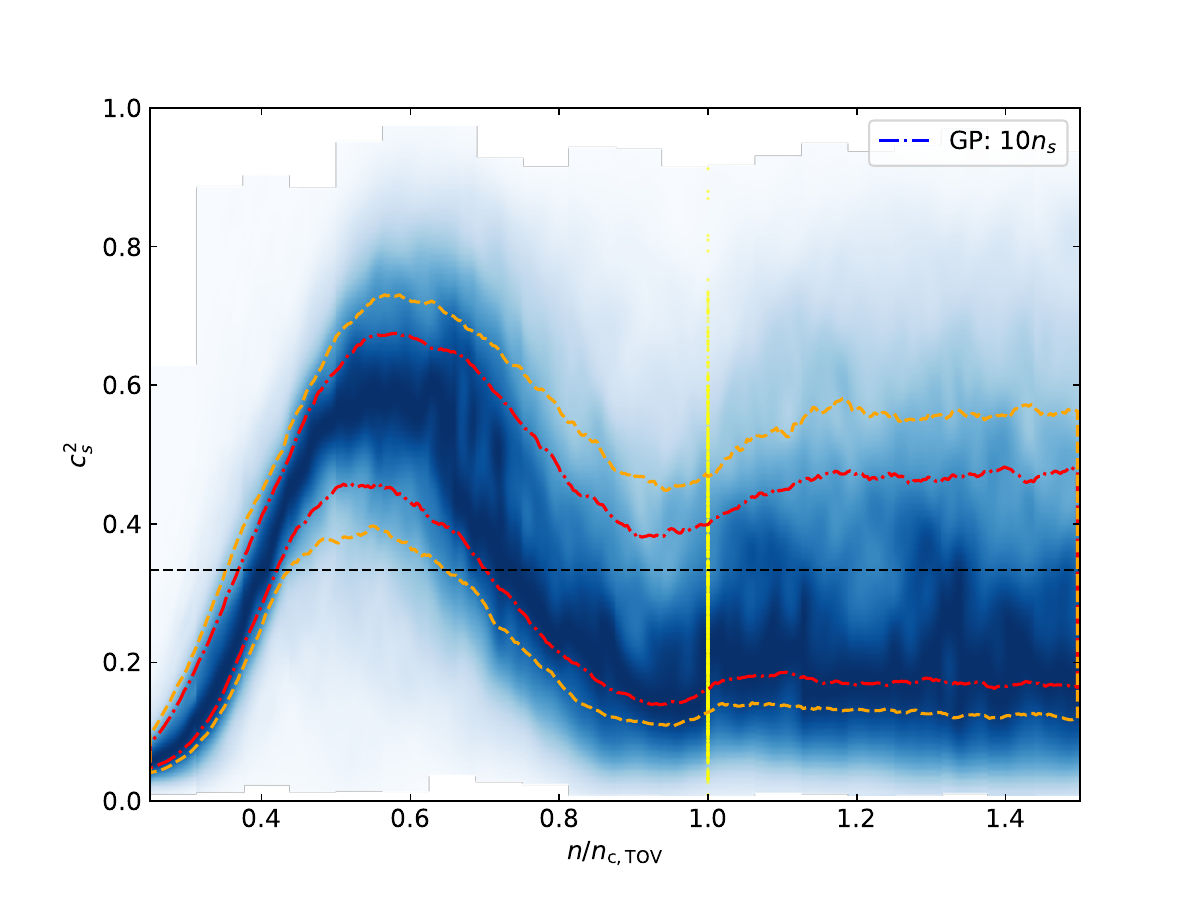}
\vspace{-0.3cm}
\caption{The top panel is the re-constructed $c_{\rm s}^{2}$ as a function of $n/n_{\rm c,TOV}$ for $n_{\rm L}=n_{\rm c,TOV}$, where the part of $n>n_{\rm c,TOV}$ is a simple extrapolation of the low density EoS. The left part of the middle panel is the same as that of the top panel, while the right part is derived by requiring the extrapolated EoS at higher densities (i.e., $\geq n_{\rm c,TOV}$) to also satisfy the pQCD likelihood at $10n_{\rm s}$ with three constant-sound-speed segment approximation (i.e., a specific PW approach). The bottom panel shows the constrained EoS with the pQCD likelihood at $n_{\rm L}=10n_{\rm s}$ for a comparison. The red dash-dotted curves and orange dashed curves denote the 50\% and 68.3\% intervals, respectively.}
\label{fig:C_S-n_l}
\end{figure}

In the above paragraph we have examined how $M_{\rm TOV}$ would be influenced by the choice of $n_{\rm L}=n_{\rm c,TOV}$ in comparison to the case of $n_{\rm L}=10n_{\rm s}$. Then what would happen beyond the density of $n_{\rm c,TOV}$?  
At densities of $n\leq n_{\rm c,TOV}$, we simply fix the EoS reconstructed with $n_{\rm L}=n_{\rm c,TOV}$, and then constrain the EoS beyond $n_{\rm c,TOV}$ with the pQCD likelihood at $10n_{\rm s}$. For simplicity, at the densities of $n\leq n_{\rm c,TOV}$ we adopt the GP method, otherwise we take the constant sound speed approach (three segments are assumed). In comparison to the case of simple extrapolation of EoSs obtained for $n_{\rm L}=n_{\rm c,TOV}$, within the density range between $n_{\rm c,TOV}$ and $10n_{\rm s}$, the $c_{\rm s}^{2}$ found in our current approach is characterized by a discontinuity, i.e., there is a sudden drop to a low value, likely indicating a strong phase transition (see the middle panel of Fig.~\ref{fig:C_S-n_l}). This is anticipated since for $n_{\rm L}=n_{\rm c,TOV}$ the $c_{\rm s}^2$ at the high densities can reach $\sim 1$ (see the top panel of Fig.~\ref{fig:C_S-n_l}), which can not smoothly connect to the pQCD constraint at $10n_{\rm s}$, in contrast with that for a direct constraint with $n_{\rm L}=10n_{\rm s}$ (see the bottom panel of Fig.~\ref{fig:C_S-n_l}). In reality, it is lack of convincing reason why a sudden transition takes place exactly at $n_{\rm c,TOV}$ rather than at higher densities. A simple but likely reasonable empirical treatment on the problem may be to set the pQCD constraint at a density uniformly distributed between $n_{\rm c,TOV}$ and $10n_{\rm s}$ (i.e, the random check). The bounds on $M_{\rm TOV}$ set by such an approach are stronger than the case of $n_{\rm L}=n_{\rm c,TOV}$ and are similar to the case of $n_{\rm L}=10n_{\rm s}$, as shown in the middle panel of Fig.~\ref{fig:M_TOV-check}. In view of the above facts, we conclude that the $M_{\rm TOV}$ value reported in this work is robust. 

As for the GP method, one of the important parameters is the correlation length. \citet{Gorda:2022jvk} has taken the mean of the correlation length to be $\bar{l}=1n_{\rm s}$, while we take $\bar{l}=0.5n_{\rm s}$ as our fiducial correlation length. To check whether the constraint on $M_{\rm TOV}$ is sensitively dependent of the correlation length, we have also reproduced the calculation with the mean of the correlation length of $\bar{l}=1n_{\rm s}$ and $\bar{l}=0.2n_{\rm s}$, respectively. As shown in the bottom panel of Fig.~\ref{fig:M_TOV-check}, the results are insensitive of the choice of $\bar{l}$.

\section{Conclusion and Discussion} 
$M_{\rm TOV}$ is one of the key parameters of the compact objects. Its value, however, can not be directly/accurately informed from the first principle and is still uncertain. Most estimates made in the literature are sensitively dependent on the model assumptions. Two approaches do not suffer from such a problem, including the cutoff modeling of the mass function of the NSs, and the EoS reconstruction with the multi-messenger data of NSs as well as the theoretical $\chi$EFT and pQCD constraints. In this work for the first time we take the advantages of these two approaches to jointly yield the precision inference of $M_{\rm TOV}=2.25^{+0.08}_{-0.07}~M_\odot$ and $R_{\rm TOV}=11.90^{+0.63}_{-0.60}$ km, with the NS mass measurement sample updated till April 2023, the reliable mass/radius data of four NSs (including PSR J0030+0451, PSR J0740+6620, and the two involved in GW170817) as well as the theoretical $\chi$EFT (pQCD) constraints at low (very high) energy densities. Such a precise value is thanks to the suppression of the high $M_{\rm TOV}$ by the pQCD constraint and the exclusion of the low $M_{\rm TOV}$ by the mass distribution information. We have adopted 
three very different EoS (non)parameterization approaches, including the single-layer feed-forward neural network model developed in \cite{Han:2022rug}, the flexible piecewise linear sound speed 
method of \cite{Jiang:2022tps}, and the Gaussian Process widely used in the literature \cite{PhysRevD.101.123007,PhysRevC.102.055803,Gorda:2022jvk}. 
The resulting $M_{\rm TOV}$ and $R_{\rm TOV}$ are found to be almost exactly the same (see Fig.~\ref{fig:M_TOV}), suggesting that our result is robust.  
 
We have also tested the case of $n_{\rm L}=n_{\rm c,TOV}$ and found a bit higher median value of $M_{\rm TOV}$ and a $\approx 0.1M_\odot$ error bar. However, such a ``conservative" approach yields a discontinuity of $c_{\rm s}^{2}$ at the density of $n_{\rm c,TOV}$ to satisfy the pQCD limit at $10n_{\rm s}$ and a more reasonable/empirical approach of $n_{\rm L} \sim {\rm Uniform}(n_{\rm c,TOV},~10n_{\rm s})$ would yield $M_{\rm TOV}$ rather similar to that of $n_{\rm L}=10n_{\rm s}$ (see Fig.~\ref{fig:M_TOV-check} for the details). 
We conclude that the parameter $M_{\rm TOV}$ is indeed precisely constrained with the current multi-messenger data of NSs and the information of the latest theoretical calculations, and the stiff EoS models with $M_{\rm TOV}\geq 2.4M_{\odot}$ is disfavored at a confidence level above 95$\%$. 
Consequently, the $2.5-3M_\odot$ compact objects detected by the LIGO/Virgo gravitational wave detectors \cite{2021arXiv211103606T} strongly suggest the presence of a group of very light black holes and hence the absence of the so-called mass gap between NSs and black holes.

One straightforward application of our results is to estimate the fate of the remnant of double NS mergers with a total mass of $m_{\rm T}$. With Eq.(7) and Eq.(8) of \cite{Shao:2020bzt}, we need $m_{\rm T}\geq M_{\rm TOV}[1.036+0.081(M_{\rm TOV}/1M_{\odot})][1-0.091(m_{\rm loss}/1M_\odot))+m_{\rm loss}\approx 2.76M_\odot$ to yield a black hole, and $m_{\rm T}< M_{\rm TOV}[0.903+0.079(M_{\rm TOV}/1M_{\odot})][1-0.091(m_{\rm loss}/1M_\odot))+m_{\rm loss}\approx 2.46M_\odot$ to result in a stable massive neutron star, otherwise a supramassive neutron star will be formed, where $m_{\rm loss}\approx 0.05M_\odot$ is the typical mass lost apart from the remnant core. For the binary NS systems summarized in Table 1, the mergers of the heaviest ones would yield black holes while the others would form supramassive NSs, in agreement with the indications of current observations of some gamma-ray bursts likely from double NS mergers \cite{2019LRR....23....1M}. The accurate estimation of $M_{\rm TOV}$ can be utilized to ascertain the nature of binary coalescence events detected by LIGO/Virgo, facilitating the differentiation between NS and BH, particularly if a mass gap exists between the two entities. Furthermore, our findings have implications for the determination of the Hubble constant, as the mass cutoff for neutron stars to be detected by GWs should align with $M_{\rm TOV}$, which is invariant with respect to redshift. The mass cutoff as inferred from the source-frame mass distribution of NSs is associated with both the redshifted mass and the redshift as predicted by the cosmological model and luminosity distance, thus offering a method to test the underlying cosmological model. The systematic investigation of this implication will be a focus of our future research.

\acknowledgments
This work is supported by the National Natural Science Foundation of China (No. 12233011), the CAS Project for Young Scientists in Basic Research (No. YSBR-088), the General Fund of the China Postdoctoral Science Foundation (No. 2023M733735 and No. 2023M733736), and the Project for Special Research Assistant of the Chinese Academy of Sciences (CAS). J.L.J. acknowledges support by the Alexander von Humboldt Foundation.

\bibliographystyle{apsrev4-2}
\bibliography{ref}

\clearpage


\newpage
\section*{appendix}

\subsection{The sample of neutron stars with a measured mass}
We summarize the up-to-date (till April 2023) measurements of the neutron star masses, totally 136 items classified into nine types. Among them, 113 items are the individual estimates of masses ($m_{\rm p}$) of neutron stars, while 17 of them are the measurements of mass function ($f$) and total mass ($m_{\rm T}$) of binary systems. The other 6 items are the measurements of mass function and mass ratio ({\it q}) of the binary systems. The details on the proper inclusion of such events in the mass function modeling can be found in \citet{Alsing:2017bbc} and \citet{Shao:2020bzt}. Please bear in mind that for some rapidly rotating pulsars, the rotation may have enhanced the gravitational mass by a factor of $0.01$ or higher. To get the reasonable mass distribution of the non-rotating objects, in our analysis we have made the corrections on the gravitational masses of the PSR J0952-0607 (spin period of 1.4ms), PSR J2215+5135 (2.61ms) and PSR J0740+6620 (2.89ms), which are $-(0.065,0.024,0.021)~M_\odot$, respectively, using the universal relation of \citet{2022ApJ...934..139K} assuming a neutron star radius of $R=12$ km.

\setlength{\tabcolsep}{2pt}
\begin{longtable*}{lcccccl}

\caption{Catalog of neutron stars with mass measurements \label{tab:mass_data}}\\
\toprule
\hline
{ID} & {Type} & {$m_{\rm p}$ ($M_\odot$)} & {$f$ ($M_\odot$)} & {$m_{\rm T}$ ($M_\odot$)} & {$q$} & {Reference}\\
\hline
\midrule
\endfirsthead

\multicolumn{7}{c}{\autoref{tab:mass_data} ({\it Continued})}\\
\toprule
\hline
{ID} & {Type} & {$m_{\rm p}$ [$M_\odot$]} & {$f$ [$M_\odot$]} & {$m_{\rm T}$ [$M_\odot$]} & {$q$} & {Reference}\\
\hline
\midrule
\endhead
\bottomrule

\hline\\
\multicolumn{7}{c}{({\it Continued on next page})}\\
\endfoot
\hline\\
\multicolumn{7}{l}{
{\bf Types:} }\\
\multicolumn{7}{l}{
NS-NS, double neutron star system; NS-WD, neutron star-white dwarf binary; NS-BH, neutron star-black hole system;}\\
\multicolumn{7}{l}{
BW, black widow millisecond pulsar system; RB, redback millisecond pulsar system; INS, isolated neutron star;}\\
\multicolumn{7}{l}{
NS-MS, neutron star-main sequence star system; HMXB, high mass x-ray binary; LMXB, low mass x-ray binary.}\\
\multicolumn{7}{l}{
The question mark means the nature of the companion is uncertain.}\\
\endlastfoot

J0453+1559 & NS-NS & 1.559$\pm$0.005 & & & & \cite{Martinez2015}\\
J0453+1559 comp. & NS-NS & 1.174$\pm$0.004 & & & & \cite{Martinez2015}\\
J0509+3801 & NS-NS & 1.34$\pm$0.08 & & & & \cite{Lynch2018}\\
J0509+3801 comp. & NS-NS & 1.46$\pm$0.08 & & & & \cite{Lynch2018}\\
J0514-4002A & NS-NS & $1.25^{+0.05}_{-0.06}$ & & & & \cite{Ridolfi2019}\\
J0514-4002A comp.& NS-NS & $1.22^{+0.06}_{-0.05}$ & & & & \cite{Ridolfi2019}\\
J0737-3039A & NS-NS & $1.338185^{+0.000012}_{-0.000014}$ & & & & \cite{Kramer2021}\\
J0737-3039B & NS-NS & $1.248868^{+0.000013}_{-0.000011}$ & & & & \cite{Kramer2021}\\
J1018-1523 & NS-NS & & 0.238062 & 2.3$\pm$0.3 & & \cite{Swiggum2022}\\
J1325-6253 & NS-NS & & 0.141517 & 2.57183$\pm$0.06 & & \cite{Sengar2022}\\
J1411+2551 & NS-NS & & 0.122390 & 2.538$\pm$0.022 & & \cite{Martinez2017}\\
J1518+4904 & NS-NS & & 0.115988 & 2.7183$\pm$0.0007 & & \cite{Janssen2008}\\
B1534+12 & NS-NS & 1.3330$\pm$0.0002 & & & & \cite{Fonseca2014}\\
B1534+12 comp. & NS-NS & 1.3455$\pm$0.0002 & & & & \cite{Fonseca2014}\\
J1756-2251 & NS-NS & 1.341$\pm$0.007 & & & & \cite{Ferdman2014}\\
J1756-2251 comp. & NS-NS & 1.230$\pm$0.007 & & & & \cite{Ferdman2014}\\
J1757-1854 & NS-NS & 1.3406$\pm$0.0005 & & & & \cite{Cameron2022}\\
J1757-1854 comp. & NS-NS & 1.3922$\pm$0.0005 & & & & \cite{Cameron2022}\\
J1759+5036 & NS-NS & & 0.081768 & 2.62$\pm$0.03 & & \cite{Agazie2021}\\
J1807-2500B & NS-NS & 1.3655$\pm$0.0021 & & & & \cite{Lynch2012}\\
J1807-2500B comp. & NS-NS & 1.2064$\pm$0.0020 & & & & \cite{Lynch2012}\\
J1811-1736 & NS-NS & & 0.128121 & 2.57$\pm$0.10 & & \cite{Corongiu2007}\\
J1829+2456 & NS-NS & 1.306$\pm$0.007 & & & & \cite{Haniewicz2021}\\
J1829+2456 comp. & NS-NS & 1.299$\pm$0.007 & & & & \cite{Haniewicz2021}\\
J1906+0746 & NS-NS & 1.291$\pm$0.011 & & & & \cite{vanLeeuwen2015}\\
J1906+0746 comp. & NS-NS & 1.322$\pm$0.011 & & & & \cite{vanLeeuwen2015}\\
J1913+1102 & NS-NS & 1.62$\pm$0.03 & & & & \cite{Ferdman2020}\\
J1913+1102 comp. & NS-NS & 1.27$\pm$0.03 & & & & \cite{Ferdman2020}\\
B1913+16 & NS-NS & 1.4398$\pm$0.0002 & & & & \cite{Weisberg2010}\\
B1913+16 comp. & NS-NS & 1.3886$\pm$0.0002 & & & & \cite{Weisberg2010}\\
J1930-1852 & NS-NS & & 0.346908 & 2.54$\pm$0.03 & & \cite{Swiggum2015}\\
J1946+2052 & NS-NS & & 0.268184 & 2.50$\pm$0.04 & & \cite{Swiggum2015}\\
B2127+11C & NS-NS & 1.358$\pm$0.010 & & & & \cite{Jacoby2006}\\
B2127+11C comp. & NS-NS & 1.354$\pm$0.010 & & & & \cite{Jacoby2006}\\
GW170817A & NS-NS & $1.47^{+0.09}_{-0.07}$ & & & & \cite{GW170817}\\
GW170817B & NS-NS & $1.27^{+0.06}_{-0.07}$ & & & & \cite{GW170817}\\
GW190425A & NS-NS & $1.56^{+0.06}_{-0.08}$ & & & & \cite{GW190425}\\
GW190425B & NS-NS & $1.74^{+0.10}_{-0.06}$ & & & & \cite{GW190425}\\
GW191219 & NS-BH & $1.17^{+0.07}_{-0.06}$ & & & & \cite{2021arXiv211103606T}\\
GW200105 & NS-BH & $1.91^{+0.33}_{-0.24}$ & & & & \cite{2021arXiv211103606T}\\
GW200115 & NS-BH & $1.44^{+0.85}_{-0.29}$ & & & & \cite{2021arXiv211103606T}\\

J0024-7204H & NS-WD(?) & & 0.001927 & 1.665$\pm$0.007 & &\cite{Freire2017}\\
J0337+1715 & NS-WD & 1.4359$\pm$0.0003 & & & & \cite{Archibald2018}\\
J0348+0432 & NS-WD & 2.01$\pm$0.04 & & & & \cite{Antoniadis2013}\\
J0437-4715 & NS-WD & 1.44$\pm$0.07 & & & & \cite{Reardon2016}\\
J0621+1002 & NS-WD & $1.53^{+0.10}_{-0.20}$ & & & & \cite{Kasian2012}\\
J0740+6620 & NS-WD & $2.072^{+0.067}_{-0.066}$ & & & & \cite{2021ApJ...918L..27R}\\
J0751+1807 & NS-WD & 1.64$\pm$0.15 & & & & \cite{Desvignes2016}\\
J0955-6150 & NS-WD & 1.71$\pm$0.02 & & & & \cite{Serylak2022}\\
J1012+5307 & NS-WD & 1.72$\pm$0.16 & & & & \cite{MataSanchez2020}\\
J1017-7156 & NS-WD & 2.0$\pm$0.8 & & & & \cite{Reardon2021}\\
J1022+1001 & NS-WD & 1.44$\pm$0.44 & & & & \cite{Reardon2021}\\
J1125-6014 & NS-WD & 1.5$\pm$0.2 & & & & \cite{Reardon2021}\\
J1141-6545 & NS-WD & 1.27$\pm$0.01 & & & & \cite{Arzoumanian2018}\\
B1516+02B & NS-WD & 2.08$\pm$0.19 & & & & \cite{Freire2008a}\\
J1528-3146 & NS-WD & $1.61^{+0.14}_{-0.13}$ & & & & \cite{Berthereau2023}\\
J1600-3053 & NS-WD & $2.3^{+0.7}_{-0.6}$ & & & & \cite{Arzoumanian2018}\\
J1614-2230 & NS-WD & 1.908$\pm$0.016 & & & & \cite{Arzoumanian2018}\\
J1713+0747 & NS-WD & 1.35$\pm$0.07 & & & & \cite{Arzoumanian2018}\\
J1738+0333 & NS-WD & $1.47^{+0.7}_{-0.6}$ & & & & \cite{Antoniadis2012}\\
J1740-5340 & NS-WD(?) & & 0.002644 & & 5.85$\pm$0.13 & \cite{Ferraro2003}\\
J1741+1351 & NS-WD & $1.14^{+0.43}_{-0.25}$ & & & & \cite{Arzoumanian2018}\\
J1748-2021B & NS-WD & & 0.000226624 & 2.69$\pm$0.071 & & \cite{Clifford2019}\\
J1748-2446am & NS-WD & $1.649^{+0.037}_{-0.11}$ & & & & \cite{Andersen2018}\\
J1748-2446an & NS-WD & & 0.02420 & 2.97$\pm$0.52 & & \cite{Ridolfi2021}\\
J1748-2446I & NS-WD & & 0.003658 & 2.17$\pm$0.02 & & \cite{Ransom2005}\\
J1748-2446J & NS-WD & & 0.013066 & 2.20$\pm$0.04 & & \cite{Ransom2005}\\
J1750-37A & NS-WD & & 0.0518649 & 1.97$\pm$0.15 & & \cite{Freire2008a}\\
B1802-07 & NS-WD & $1.26^{+0.08}_{-0.17}$ & & & & \cite{Thorsett1999}\\
J1802-2124 & NS-WD & 1.24$\pm$0.11 & & & & \cite{Ferdman2010}\\
J1811-2405 & NS-WD & $2.0^{+0.8}_{-0.5}$ & & & & \cite{Ng2020}\\
J1823-3021G & NS-WD(?) & & 0.0123 & 2.65$\pm$0.07 & & \cite{Ridolfi2021}\\
J1824-2452C & NS-WD(?) & & 0.006553 & 1.616$\pm$0.007 & & \cite{Freire2008b}\\
B1855+09 & NS-WD & $1.37^{+0.13}_{-0.10}$ & & & & \cite{Arzoumanian2018}\\
J1909-3744 & NS-WD & 1.492$\pm$0.014 & & & & \cite{Liu2020}\\
J1911-5958A & NS-WD & 1.34$\pm$0.08 & & & & \cite{Bassa2006}\\
J1918-0642 & NS-WD & 1.29$\pm$0.1 & & & & \cite{Arzoumanian2018}\\
J1946+3417 & NS-WD & 1.828$\pm$0.022 & & & & \cite{Barr2017}\\
J1949+3106 & NS-WD & $1.34^{+0.17}_{-0.15}$ & & & & \cite{Zhu2019a}\\
J1950+2414 & NS-WD & 1.496$\pm$0.023 & & & & \cite{Zhu2019a}\\
J2043+1711 & NS-WD & $1.38^{+0.12}_{-0.13}$ & & & & \cite{Arzoumanian2018}\\
J2045+3633 & NS-WD & 1.251$\pm$0.021 & & & & \cite{Mckee2020}\\
J2053+4650 & NS-WD & $1.40^{+0.21}_{-0.18}$ & & & & \cite{Berezina2017}\\
J2140-2311B & NS-WD(?) & & 0.2067 & 2.53$\pm$0.08 & & \cite{Balakrishnan2023}\\
J2222-0137 & NS-WD & 1.831$\pm$0.010 & & & & \cite{Guo2021}\\
J2234+0611 & NS-WD & $1.353^{+0.014}_{-0.017}$ & & & & \cite{Stovall2019}\\
B2303+46 & NS-WD & $1.3^{+0.13}_{-0.46}$ & & & & \cite{Thorsett1999}\\
J0952-0607 & BW & 2.35$\pm$0.17 & & & & \cite{Romani2022}\\
J1301+0833 & BW & $1.6^{+0.22}_{-0.25}$ & & & & \cite{Kandel2022}\\
J1311-3430 & BW & 2.22$\pm$0.1 & & & & \cite{Kandel2022}\\
J1555-2908 & BW & $1.67^{+0.07}_{-0.05}$ & & & & \cite{Kennedy2022}\\
J1653-0158 & BW & 2.15$\pm$0.16 & & & & \cite{Kandel2022}\\
J1810+1744 & BW & 2.11$\pm$0.04 & & & & \cite{Kandel2022}\\
J1959+2048 & BW & 1.81$\pm$0.07 & & & & \cite{Clark2023}\\
3FGL J0212.1+5320 & RB & $1.85^{+0.32}_{-0.26}$ & & & & \cite{Shahbaz2017}\\
3FGL J0427.9-6704 & RB & $1.86^{+0.11}_{-0.10}$ & & & & \cite{Strader2016}\\
2FGL J0846.0+2820 & RB & 1.96$\pm$0.41 & & & & \cite{Swihart2017}\\
J1023+0038 & RB & $1.65^{+0.19}_{-0.16}$ & & & & \cite{Strader2019}\\
J1048+2339 & RB & & 0.01 & & 5.15$\pm$0.19 & \cite{Strader2019}\\
1FGL J1417.7-4407 & RB & $1.62^{+0.43}_{-0.17}$ & & & & \cite{Swihart2018}\\
J1431-4715 & RB & & 0.000885 & & 10.42$\pm$0.11 & \cite{Strader2019}\\
J1622-0315 & RB & & 0.000436 & & 14.29$\pm$0.20 & \cite{Strader2019}\\
J1628-3205 & RB & & 0.00171 & & 8.33$\pm$0.21 & \cite{Strader2019}\\
J1723-2837 & RB & $1.22^{+0.26}_{-0.2}$ & & & &  \cite{Strader:2018qbi}\\
J1816+4510 & RB & & 0.0017607 & & 9.54$\pm$0.21 & \cite{Kaplan2013}\\
4FGL J2039.5-5617 & RB & $1.3^{+0.155}_{-0.10}$ & & & & \cite{Clark2021}\\
3FGL J2039.6-5618 & RB & $2.04^{+0.37}_{-0.25}$ & & & & \cite{Strader2019}\\
J2129-0429 & RB & 1.74$\pm$0.18 & & & & \cite{Bellm2016}\\
J2215+5135 & RB & $2.28^{+0.10}_{-0.09}$ & & & & \cite{Kandel2020}\\
J2339-0533 & RB & 1.47$\pm$0.09 & & & & \cite{Kandel2020}\\

J0030+0451 & INS & $1.34^{+0.15}_{-0.16}$ & & & & \cite{2019ApJ...887L..21R}\\
J0045-7319 & NS-MS & 1.58$\pm$0.34 & & & & \cite{Nice2003}\\
J1903+0327 & NS-MS & $1.666^{+0.010}_{-0.012}$ & & & & \cite{Arzoumanian2018}\\
4U1538-522 & HMXB & 1.02$\pm$0.17 & & & & \cite{Falanga2015}\\
4U1700-377 & HMXB & 1.96$\pm$0.19 & & & & \cite{Falanga2015}\\
Cen X-3 & HMXB & 1.57$\pm$0.16 & & & & \cite{Falanga2015}\\
EXO 1722-363 & HMXB & 1.91$\pm$0.45 & & & & \cite{Falanga2015}\\
Her X-1 & HMXB & 1.07$\pm$0.36 & & & & \cite{Rawls2011}\\
J013236.7+303228 & HMXB & 2.0$\pm$0.4 & & & & \cite{Bhalerao2012}\\
LMC X-4 & HMXB & 1.57$\pm$0.11 & & & & \cite{Falanga2015}\\
OAO 1657-415 & HMXB & 1.74$\pm$0.3 & & & & \cite{Falanga2015}\\
SAX J1802.7-2017 & HMXB & 1.57$\pm$0.25 & & & & \cite{Falanga2015}\\
SMC X-1 & HMXB & 1.21$\pm$0.12 & & & & \cite{Falanga2015}\\
Vela X-1 & HMXB & 2.12$\pm$0.16 & & & & \cite{Falanga2015}\\
XTE J1855-026 & HMXB & 1.41$\pm$0.24 & & & & \cite{Falanga2015}\\
2S 0921-630 & LMXB & 1.44$\pm$0.1 & & & & \cite{Steeghs2007}\\
4U 1608-52 & LMXB & $1.57^{+0.30}_{-0.29}$ & & & & \cite{Ozel2016}\\
4U1702-429 & LMXB & 1.9$\pm$0.3 & & & & \cite{Nattila2017}\\
4U 1724-207 & LMXB & $1.81^{+0.25}_{-0.37}$ & & & & \cite{Ozel2016}\\
4U 1820-30 & LMXB & $1.77^{+0.25}_{-0.28}$ & & & & \cite{Ozel2016}\\
Cyg X-2 & LMXB & 1.71$\pm$0.21 & & & & \cite{Casares2010}\\
KS 1731-260 & LMXB & $1.61^{+0.35}_{-0.37}$ & & & & \cite{Ozel2016}\\
EXO 1745-248 & LMXB & $1.65^{+0.21}_{-0.31}$ & & & & \cite{Ozel2016}\\
SAX J1748.9-2021 & LMXB & $1.81^{+0.25}_{-0.37}$ & & & & \cite{Ozel2016}\\
X 1822-371 & LMXB & 1.96$\pm$0.36 & & & & \cite{Munoz2005}\\
XTE J2123-058 & LMXB & 1.53$\pm$0.42 & & & & \cite{Gelino2002}\\

\end{longtable*}

\subsection{The impact of different mass distribution models and NS sample}
\begin{figure*}[htbp]
\centering
\vspace{-0.3cm}
\includegraphics[width=0.49\textwidth]{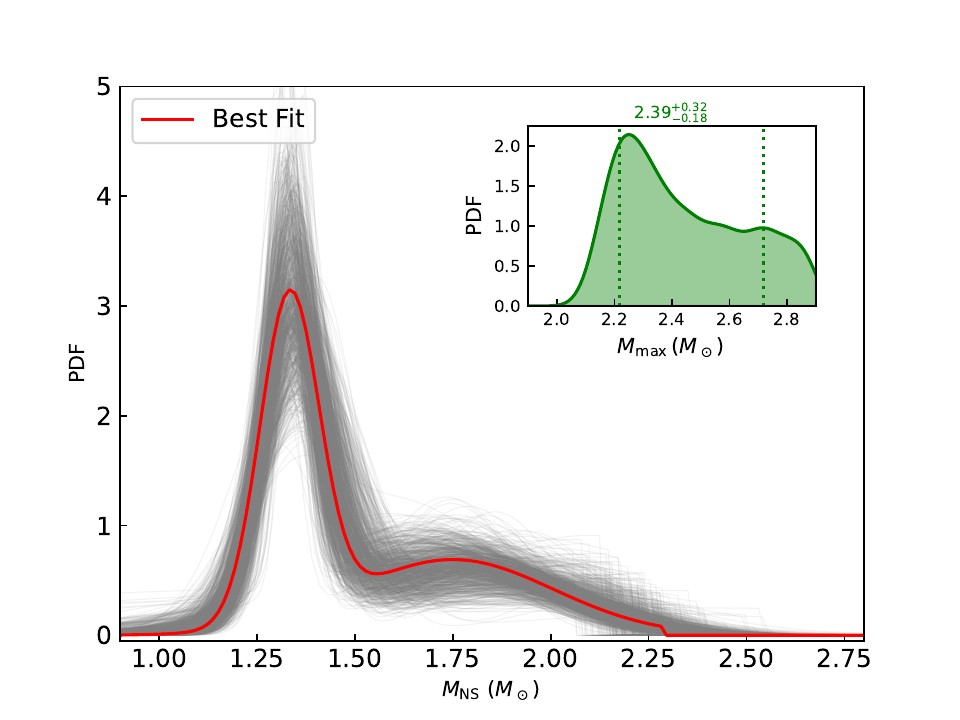}
\includegraphics[width=0.49\textwidth]{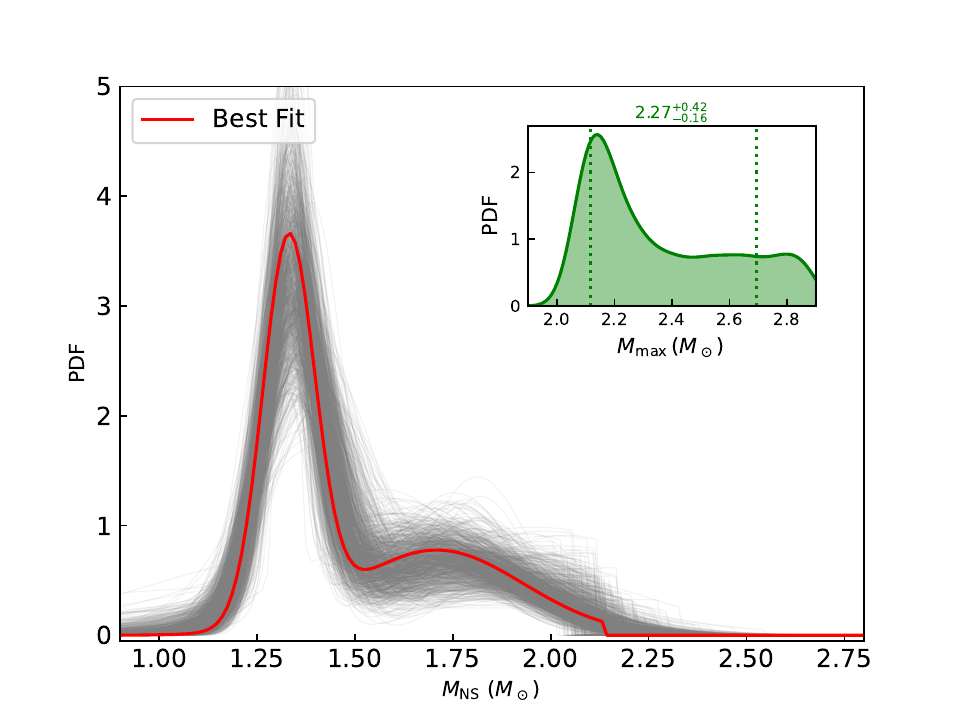}
\includegraphics[width=0.49\textwidth]{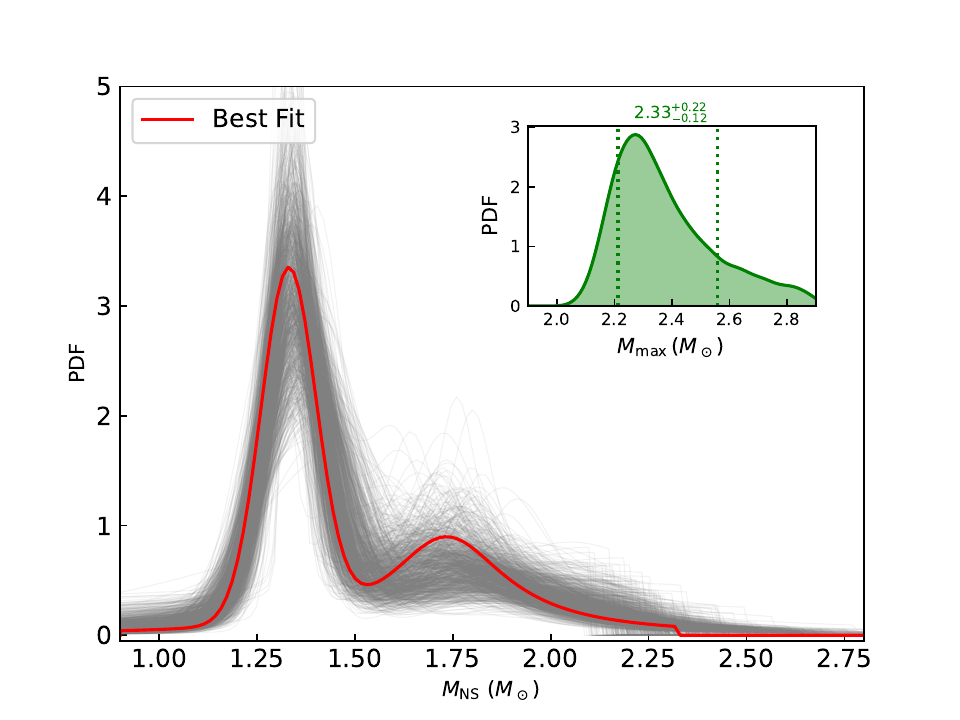}
\includegraphics[width=0.49\textwidth]{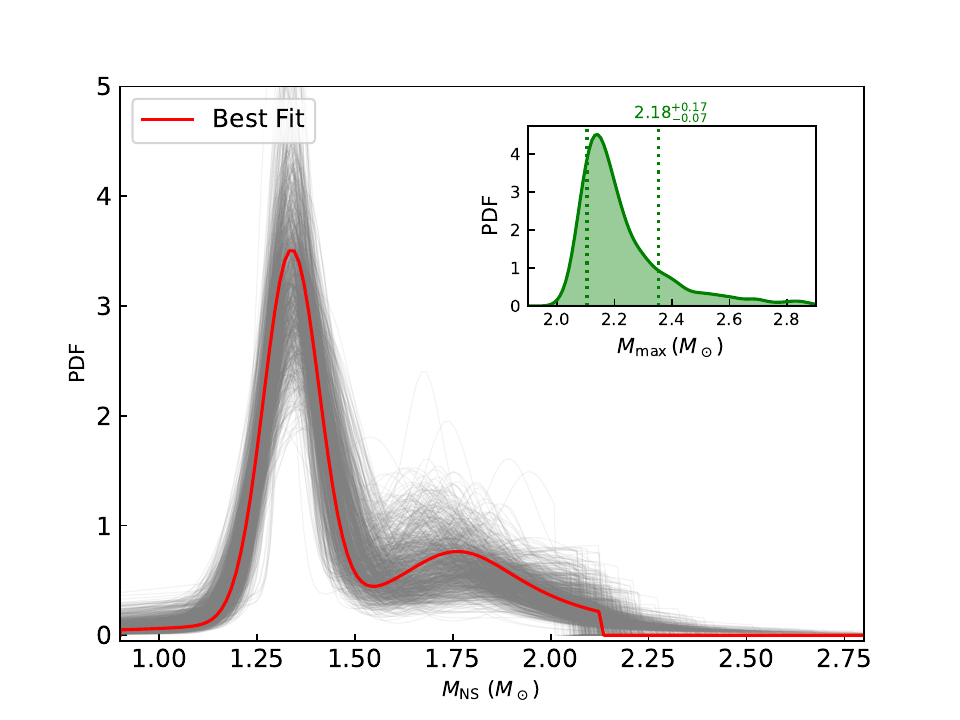}
\vspace{-0.3cm}
\caption{Similar to Fig.~\ref{fig:mass_dist}. The top row presents the outcomes derived from the two-component Gaussian mixture model, whereas the bottom row corresponds to the results utilizing the Gaussian+Cauchy-Lorentz model. The left column excludes the data from GW170817, PSR J0030+0451, and PSR J0740+6620, whereas the right column presents the results obtained after omitting some high-mass NSs.}
\label{fig:mass_dist_test}
\end{figure*}

\begin{figure*}[htbp]
\centering
\vspace{-0.3cm}
\includegraphics[width=0.98\textwidth]{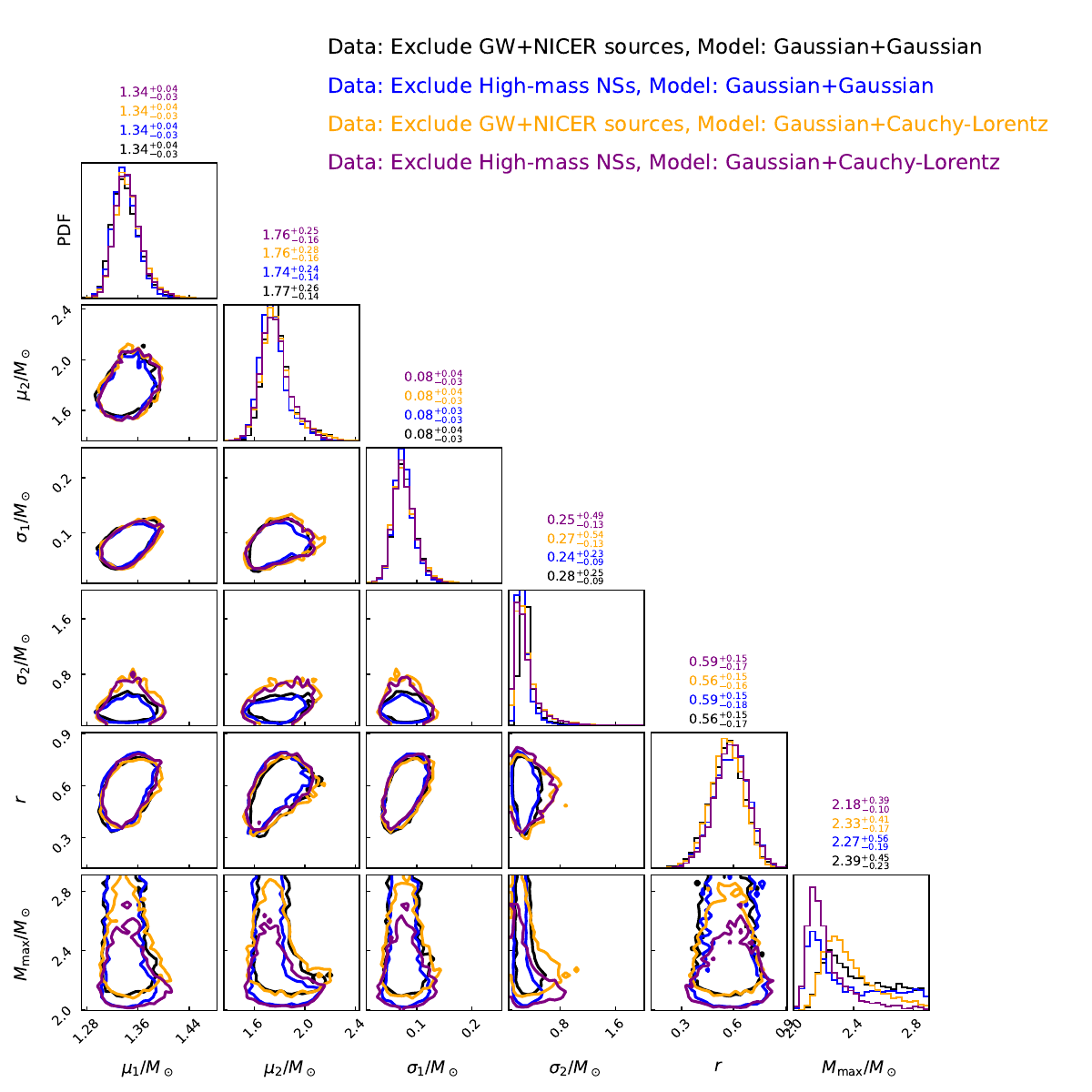}
\vspace{-0.3cm}
\caption{Posterior distribution of mass distribution parameters derived from various tests. The parameters $\mu_1$ and $\sigma_1$ represent the location and scale, respectively, for the first component, whereas $\mu_2$ and $\sigma_2$ correspond to those for the second component. The variable $r$ denotes the weight attributed to the first component.}
\label{fig:corner}
\end{figure*}

\begin{figure*}[htbp]
\centering
\vspace{-0.3cm}
\includegraphics[width=0.98\textwidth]{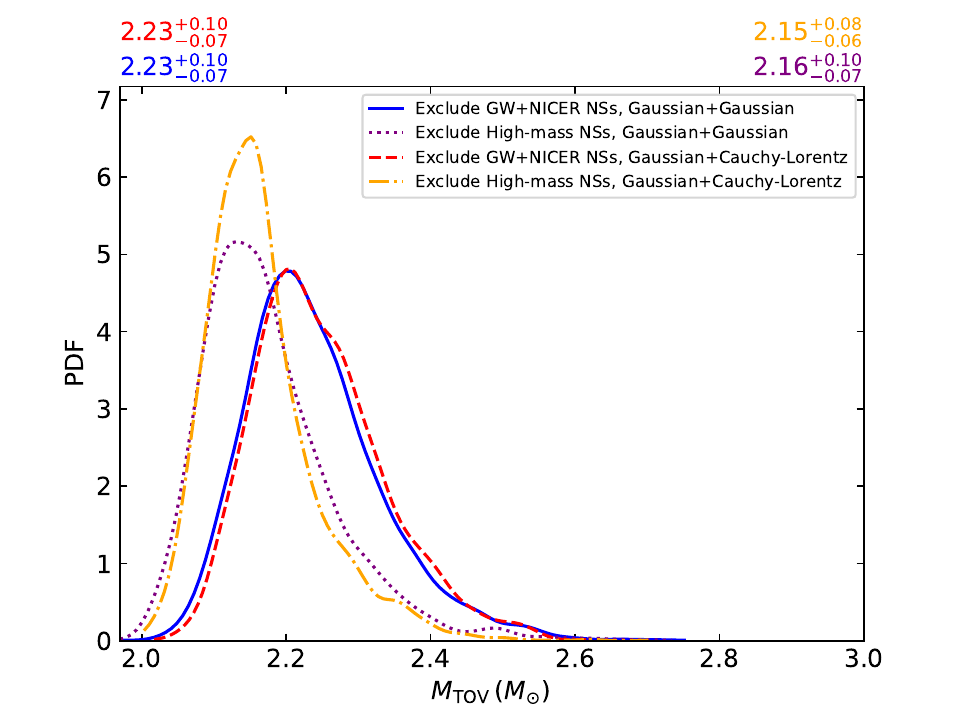}
\vspace{-0.3cm}
\caption{Posterior distributions of $M_{\rm TOV}$ derived from different $P(M_{\rm max})$ results, based on the GP EoS reconstruction approach in the case of $n_{\rm L}=10n_s$.}
\label{fig:mtov_gp_test}
\end{figure*}

In this subsection, we detail the results of the NS mass distribution derived from various models and datasets. The code for reproducing these findings is available on the \href{https://github.com/ShaoPengTang/ShowMyWorks/blob/main/calc_pmmax.py}{GitHub page}.
Initially, we modified the bimodal Gaussian mixture model by substituting its second component with a Cauchy-Lorentz distribution. Additionally, we investigated the impact of omitting certain NS sources on the robustness of our results by excluding specific datasets.
The first group of sources omitted includes GW170817, PSR J0030+0451, and PSR J0740+6620. While the second group of omitted sources consists of certain NSs (i.e., PSR J2215+5135, PSR J0952-0607, PSR J1311-3430, and PSR J1748-2021B) characterized by high mass and relatively smaller measurement errors, a combination anticipated to exert more impact on the determination of $M_{\rm max}$.
As depicted in Fig.~\ref{fig:mass_dist_test} and Fig.~\ref{fig:corner}, our analyses reveal that different models yield consistent parameter distributions. However, the removal of high-mass NSs results in a lower maximum mass ($M_{\rm max}$) estimate, as expected. Furthermore, with respect to the impact on the $M_{\rm TOV}$, we illustrate using the GP method as an example. As shown in Fig.~\ref{fig:mtov_gp_test}, we find that the main findings reported in the main text are demonstrated to be substantially robust against the choice of models.

\clearpage

\end{document}

%% file: aas_macros.tex
\def\apj{ApJ~}                 
\def\apjl{ApJL~}                
\def\apjs{ApJS~}               
\def\aap{A\&A~}                

\def\mnras{MNRAS~}             
\def\prd{Phys~Rev~D~}        
\def\prl{Phys~Rev~Lett~}    
\def\nat{Nature~}              

